\documentstyle[amsmath,amssymb,graphicx]{article}

\makeatletter
\newdimen\normalarrayskip              
\newdimen\minarrayskip                 
\normalarrayskip\baselineskip
\minarrayskip\jot
\newif\ifold             \oldtrue            \def\new{\oldfalse}
\def\arraymode{\ifold\relax\else\displaystyle\fi} 
\def\eqnumphantom{\phantom{(\theequation)}}     
\def\@arrayskip{\ifold\baselineskip\z@\lineskip\z@
     \else
     \baselineskip\minarrayskip\lineskip2\minarrayskip\fi}
\def\@arrayclassz{\ifcase \@lastchclass \@acolampacol \or
\@ampacol \or \or \or \@addamp \or
   \@acolampacol \or \@firstampfalse \@acol \fi
\edef\@preamble{\@preamble
  \ifcase \@chnum
     \hfil$\relax\arraymode\@sharp$\hfil
     \or $\relax\arraymode\@sharp$\hfil
     \or \hfil$\relax\arraymode\@sharp$\fi}}
\def\@array[#1]#2{\setbox\@arstrutbox=\hbox{\vrule
     height\arraystretch \ht\strutbox
     depth\arraystretch \dp\strutbox
     width\z@}\@mkpream{#2}\edef\@preamble{\halign
\noexpand\@halignto
\bgroup \tabskip\z@ \@arstrut \@preamble \tabskip\z@ \cr}%
\let\@startpbox\@@startpbox \let\@endpbox\@@endpbox
  \if #1t\vtop \else \if#1b\vbox \else \vcenter \fi\fi
  \bgroup \let\par\relax
  \let\@sharp##\let\protect\relax
  \@arrayskip\@preamble}
\def\eqnarray{\stepcounter{equation}%
              \let\@currentlabel=\theequation
              \global\@eqnswtrue
              \global\@eqcnt\z@
              \tabskip\@centering
              \let\\=\@eqncr

 \halign to \displaywidth\bgroup
    \eqnumphantom\@eqnsel\hskip\@centering
    $\displaystyle \tabskip\z@ {##}$%
    \global\@eqcnt\@ne \hskip 2\arraycolsep
         $\displaystyle\arraymode{##}$\hfil
    \global\@eqcnt\tw@ \hskip 2\arraycolsep
         $\displaystyle\tabskip\z@{##}$\hfil
         \tabskip\@centering
    &{##}\tabskip\z@\cr}
\newfont{\hr}{msbm10}
\newfont{\ams}{msam10}

\def\be{\begin{eqnarray}}
\def\ee{\end{eqnarray}}
\def\nn{\nonumber}

\def\beq{\begin{equation}}
\def\eeq{\end{equation}}
\def\ba{\beq\new\begin{array}{c}}
\def\ea{\end{array}\eeq}
\def\be{\ba}
\def\ee{\ea}

\def\p{\partial}

\def\r{\rho}

\hoffset= - 1.0in         

\title{{\bf {Towards matrix model representation of HOMFLY polynomials} \vspace{.2cm}}
\author{{\bf A.Alexandrov}\footnote{ {\small {\it Freiburg Institute for Advanced Studies (FRIAS), University of Freiburg, Germany  and Mathematics Institute, University of Freiburg, Germany} and {\it ITEP, Moscow, Russia}}; alexandrovsash@gmail.com
}, {\bf A.Mironov}\footnote{ {\small {\it
Lebedev Physics Institute} and {\it ITEP, Moscow, Russia}};
mironov@itep.ru; mironov@lpi.ru}, {\bf A.Morozov}\thanks{{\small
{\it ITEP, Moscow, Russia}}; morozov@itep.ru} \ and {\bf And.Morozov}\thanks{{\small {\it
Moscow State University} and {\it ITEP, Moscow, Russia} and {\it Laboratory of Quantum Topology,
Chelyabinsk State University, Chelyabinsk, Russia}};
Andrey.Morozov@itep.ru}}\date{ }}

\begin{document}

\maketitle

\vspace{-6.0cm}

\begin{center}
\hfill FIAN/TD-07/14\\
\hfill ITEP/TH-11/14\\
\end{center}

\vspace{4.5cm}

\begin{abstract}
We investigate possibilities of generalizing the TBEM eigenvalue
matrix model, which represents
the non-normalized colored HOMFLY polynomials for torus knots
as averages of the corresponding characters.
We look for a model of the same type, which
is a usual Chern-Simons mixture of the Gaussian potential,
typical for Hermitean models, and the sine Vandermonde factors, typical
for the unitary ones.
We mostly concentrate on the family of twist knots,
which contains a single torus knot, the trefoil.
It turns out that for the trefoil the TBEM measure
is provided by an action of Laplace exponential
on the Jones polynomial.
This procedure can be applied to arbitrary knots
and provides a TBEM-like integral representation
for the $N=2$ case.
However, beyond the torus family,
both the measure and its lifting to larger $N$
contain non-trivial corrections in $\hbar=\log q$.
A possibility could be to absorb these corrections
into a deformation of the Laplace evolution by higher
Casimir and/or cut-and-join operators,
in the spirit of Hurwitz $\tau$-function approach to knot theory,
but this remains a subject for future investigation.
\end{abstract}

\subsection{Introduction}

Knot polynomials \cite{knpols} are examples of the Hurwitz $\tau$-function \cite{AMMN},
a new and intriguing generalization of the free-fermion \cite{Japs} KP/Toda
$\tau$-functions, probably related to non-Abelian $\tau$-functions of \cite{GKLMM}.
As such they should possess a number of different realizations:
as functional integrals in free field and topological theories \cite{UFN2},
as matrix models of the ordinary and Kontsevich types \cite{UFN3},
as various $W$-representations \cite{MMS} a la \cite{MorSha,Al,AMMNlast}.
While the first of these representations is well known:
knot polynomials are Wilson line averages in Chern-Simons theory \cite{CS,Wit}
and/or results of $R$-matrix (modular group) evolution
of conformal blocks \cite{Wit,Ind,RT},
all the other realizations are more-or-less available only for the very
specific class of torus knots and links:
this story is mostly around the Rosso-Jones formula \cite{RJ}.
In particular, the matrix model representation is known only for
the unknot (Chern-Simons partition function) \cite{ZCS}
\be
Z_{CS} = \int   \prod_{i<j}^N \sinh^2(u_i-u_j)
\prod_{i=1}^N \exp\left(-\frac{u_i^2}{2\hbar} \right)du_i
\label{Zcs}
\ee
and for arbitrary $[m,n]$ torus link/knot \cite{Tierz,BEM}:
\be
H_R^{[m,n]}(q|A)\Big|_{q=e^\hbar,A=q^N} \sim
\int \chi_R(e^U)
\prod_{i<j}^N \sinh\left(\frac{u_i-u_j}{m}\right)\sinh\left(\frac{u_i-u_j}{n}\right)
\prod_{i=1}^N \exp\left(-\frac{u_i^2}{mn\hbar} \right)du_i
\label{BEM}
\ee
(here and everywhere in this paper knot polynomials are non-normalized).

However, despite being now available only for torus knots,
all such realizations should exist for an arbitrary family of knots,
what is strongly supported by the overwhelming success of the evolution
method \cite{DMMSS,evo}.
Still, it is a long-standing problem to generalize (\ref{BEM}), to begin with,
beyond the very special family of torus knots.
This is the goal of the present letter to make a step towards this generalization. Though the final answer remains not yet reached,
we realize a few essential properties of a possible final answer for the knot matrix model.

\bigskip

\subsection{Summary}

We are looking for an answer for the HOMFLY polynomial in the matrix model form
\be
H_R^{{\cal K}} \stackrel{?}{\sim}
{\int \chi_R(e^U)
\prod_{i<j}^N \mu_\gamma^{{\cal K}}(u_i-u_j|\tau)\sinh\left(u_i-u_j\right)
\prod_{i=1}^N \exp\left(-\frac{u_i^2}{\gamma\hbar} \right)du_i\over
\int\prod_{i<j}^N \mu_\gamma^{{\cal K}}\sinh\left(u_i-u_j|\tau\right)
\prod_{i=1}^N \exp\left(-\frac{u_i^2}{\gamma\hbar} \right)du_i}
\label{ga}
\ee
where $\sim$ means a factor that depends on $q$, $N$
and representation $R$ in a controllable way.
$\gamma$ is a yet unknown constant,
and we choose an anzatz for the measure to depend on $N$ only through a function $\tau(N,\hbar)$.
We propose to construct such generalization of the TBEM model \cite{Tierz,BEM}
to the twist knots in a few steps.
\begin{itemize}
\item
First, one considers the case of $N=2$: then the question is,
what is the relevant integral representation of the Jones polynomial.
The answer is universal:
since the inverse of the integral transform
\be
{\cal G}(\rho) = e^{-\frac{\gamma\rho^2}{2\hbar}}\int_{-\infty}^\infty
\mu(u)\cdot\sinh\left(\frac{\rho u}{\hbar}\right)  \exp\left(-\frac{u^2}{2\gamma\hbar}\right)du
\label{intra}
\ee
(we need it in application to odd functions)
is
\be
\mu(u) = e^{-\frac{\gamma\hbar}{2}\p^2_u} {\cal G}\left(\frac{u}{\gamma}\right)
\ee
the measure in the matrix-model integral is made from
the Laplace evolution of the Jones polynomial ${\cal J}^{\cal K}(\rho|\hbar)$
for the
knot ${\cal K}$,
rewritten in appropriate variables $(\rho,\hbar)$. After making a substitution $\rho=\frac{u}{\gamma}$, it is
\be
\boxed{
\mu_\gamma^{\cal K}(u|\hbar) = e^{-\frac{\gamma\hbar}{2}\p^2_u}
{\cal J}^{\cal K}\left(\frac{u}{\gamma}\,\Big|\,\hbar\right)
}
\label{LevJones}
\ee
This, however, gives the answer without dividing by the normalization integral as in (\ref{ga}),
which leads to the normalization factor in (\ref{ga}).
Note that for Jones polynomials the role of $\tau$ is played by $\hbar$.
\item
Usually the Jones polynomial $J_{r-1}(q)$ is a function of $q$ and of the spin $(r-1)/2$ of representation of $SU(2)$. Eq.(\ref{LevJones})
deals with ${\cal J}(\rho|\hbar)=J_{\rho/\hbar-1}(e^\hbar)$ obtained from $J_{r-1}(q)$
by the substitution $\  r\ \longrightarrow\ \r=r\hbar\ $,
well familiar from the study of Kashaev limit \cite{Kash} and Hikami invariants \cite{Hik}, and we denote the function
of these new variables $(\r,\hbar)$ by the calligraphic letter, ${\cal J}(\rho|\hbar)$ which implies some analytic continuation in the discrete index $r$ described below. In fact, since we will be not able to perform an exact integration in (\ref{ga}), we study series in $\hbar$.
\item
The next step could be equally universal:
the $N$-fold integral for HOMFLY polynomial is made from this measure by
direct analogue of (\ref{BEM}):
\vspace{-0.3cm}
\be
H_R^{\cal K}  = \frac{\Big<\chi_R\Big>^{\cal K}}{\Big< 1\Big>^{\cal K}}
\label{Hasav}
\ee
where
\be
\boxed{
\Big< G\Big>^{\cal K} \stackrel{?}{=}
\int  G(e^U) \prod_{i<j}^N \Big(\mu_\gamma^{\cal K}(u_i-u_j|\tau)\cdot \sinh(u_i-u_j)\Big)
\prod_{i=1}^N \exp\left(-\frac{u_i^2}{\gamma\hbar} \right)du_i
}
\label{aveL}
\ee
and $\mu_\gamma^{\cal K}(u|\tau)$ is an {\it odd} function of the integration variable $u$.
Note that $\tau$ substitutes $\hbar$ in the both places in (\ref{LevJones}): in the Laplace evolution and in ${\cal J}^{\cal K}$.
\item
However, it is of course impossible to reconstruct HOMFLY polynomial from Jones
in a universal way: something in this reconstruction should depend on the {\it type}
of the knot.
For the two simplest families, of torus and twist knots the difference is basically
in the choice of the evolution parameter $\tau$:
\be\label{tau}
{\rm for\ torus\ links/knots} \ \ \  \tau = \hbar = \log q  \\
{\rm for\ twist\ knots} \ \ \  \tau = \frac{1}{2}N\hbar = \log A^{1/2}
\ee
what is in perfect accordance with what we know from the study evolution method in \cite{evo}.

It is an intriguing question, what happens for other families.
But now the way is open to study this kind of problems --
which look very promising.
\item Even for the twist knots (\ref{aveL}) at $N\neq 2$
holds only up to the order $\hbar^5$ and needs to be corrected, see s.5.6.
\end{itemize}

\bigskip

\subsection{The role of $\gamma$}

What happens in the case of {\it torus} links/knots, is that there is an additional
great simplification: one can choose auxiliary parameter $\gamma$
in such a way, that the result of Laplace evolution in (\ref{LevJones})
gets $\hbar$-independent and actually the measure gets nearly trivial --
namely, reduces to that in (\ref{BEM}).
The choice is clear from (\ref{BEM})
\be
\gamma^{[m,n]} = -mn
\ee
(note the sign minus indicating a non-naive choice of integration contour,
or analytical continuation of the answer, if one prefers, which implicit in (\ref{BEM}).

It is an open question for us, what is the meaning of this
spectacular possibility, and if some counterpart of it exists in general.
Even for the twist knots we have not yet resolved this problem.

\bigskip

Now we provide some evidence in support of above claims.
We discuss the family of twist knots, following the
description in \cite[s.5.2]{evo}, which we
assume the reader to be familiar with.
In this brief presentation all the torus links/knots
will be represented by a single trefoil, which is also
a member of the twist family.
All the claims, illustrated by this is example, are actually
true for entire torus family.

\bigskip

\subsection{Jones polynomials}

According to general principles of the link differential calculus
\cite{DGR,GGS,Arth,AnoMMM21}, the HOMFLY polynomial is decomposed into a sum of
products of the quantities $\{Aq^a\} = Aq^a - A^{-1}q^{-a}$.
In particular, for the  Jones polynomial there is usually a hypergeometric type
expansion \cite{Gar}, which is especially nice
for unreduced Jones:
\be
\{q\}J_{[r-1]}^{\cal K} (q) = \sum_{s=0}^{r-1} F_s^{\cal K}  \prod_{j=-s}^s \{q^{r+j}\}=
\sum_{s=0}^{r-1}  2^{2s+1}F_s^{\cal K} \prod_{j=-s}^s \sinh (r\hbar+j\hbar)=
\\= \{q\}[r]+\{q\}^3[r-1][r][r+1]\cdot F_1^{\cal K} + \{q\}^5[r-2][r-1][r][r+1][r+2]\cdot F_2^{\cal K}
+ \ldots
\label{Jr}
\ee
where the square brackets denote quantum numbers $[r] =\{q^r\}/\{q\}$.
Note that we shifted the labeling of representations by one to simplify the formulas below.
The coefficient functions $F_s$ are polynomials in $A=q^N=e^{N\hbar}$ and $q=e^{\hbar}$,
in (\ref{Jr}) they are reduced to $N=2$.
These functions are especially simple for the twist knots \cite{Ito41,evo}.
For the twist knot number $k$ we get from sec.5.2 of \cite{evo}:
\be
F_s^{(k)} = q^{s(s-1)/2}A^s \sum_{j=0}^s \frac{(-)^j[s]!}{[j]![s-j]!}\frac{(Aq^{j-1})^{2jk}\{Aq^{2j-1}\}}
{\prod_{i=j-1}^{s+j-1}\{Aq^i\}}=(-k)^s+O(\hbar)
\label{Fsk}
\ee
The figure eight knot $4_1$ corresponds to $k=-1$, the trefoil $3_1$ to $k=1$,
unknot arises at $k=0$. The Rolfsen table notation \cite{katlas} is $(2|k|+2)_1$ for negative $k$
and $(2k+1)_2$ for positive $k$ (for $3_1$ the Rolfsen labeling is not smooth:
$3_2$ is actually $3_1$, since it is the only knot with plane projection having only three intersections).

Since
\be
\prod_{j=-s}^s \{q^{r+j}\} =  \frac{1}{2}\sum_{j=0}^{2s+1} (-)^j\frac{[2s+1]!}{[j]![2s+1-j]!}
\left\{q^{(2s+1-2j)r}\right\}
\ \  \Longrightarrow   \\
\sum_{s=0}^M F_s \cdot \prod_{j=-s}^s \{q^{r+j}\} =
\sum_{p=0}^M \sum_{j=0}^{M-p} (-)^j\frac{[2p+2j+1]!}{[j]![2p+j+1]!}\left\{q^{(2p+1)r}\right\} \cdot F_{p+j}
\ee
We shifted $s=p+j$ at the r.h.s., and the last relation is true
for any  $M$ (which can thus be put equal to infinity).
Further, since
\be
q^{-\frac{\gamma}{2}\partial_u^2} \left\{q^{(2p+1)\frac{u}{\gamma\bar h}}\right\} =
2q^{-\frac{(2p+1)^2}{2\gamma}} \sinh\left( \frac{(2p+1)u}{\gamma}\right)
\nn
\ee
one gets a hypergeometric-like representation for the measure (\ref{LevJones})
\be
\boxed{
\mu_\gamma^{{\cal K}}(u|\hbar)=
\sum_{p=0}^\infty q^{-(2p+1)^2/2\gamma}
\sinh\left(\frac{(2p+1)u}{\gamma}\right)
\sum_{j=0}^\infty  (-)^j \frac{[2p+2j+1]!}{[j]![2p+j+1]!}\!\left.F_{p+j}^{\cal K} \right|_{A=q^2}
}
\label{mesdoubleser}
\ee

\subsection{Comments}

\subsubsection{Jones polynomials in variables $(\r,\hbar)$}

Formulas similar to (\ref{mesdoubleser}) are not that simple to deal with. When performing checks for the matrix model, we
used the $\hbar$-expansion instead. Let us see how these checks are done.

One of the possibilities is to use (\ref{mesdoubleser}) directly. Say, in the leading order
$\left.F^{(k)}_s\right|_{A=q^2} = (-k)^s+ O(\hbar)$, and
\be
\!\!\!\!\sum_{j=0}^\infty  (-)^j \frac{(2p+2j+1)!}{j!(2p+j+1)!}(-k)^{p+j}=(-k)^p
\cdot\phantom{.}_2F_1(p+1,p+3/2;2p+2;4k)={2\cdot (-4k)^p\over
\sqrt{1-4k}\ \Big(1+\sqrt{1-4k}\Big)^{2p+1}}
\ee
and the sum in (\ref{mesdoubleser}) is easily calculated:
\be
\boxed{
\mu_{(0)}(u) =
\frac{\sinh\left(\frac{u}{\gamma}\right)}{1+4k\sinh^2\left(\frac{u}{\gamma}\right)}}
\ee

Another, simpler possibility is to use formula (\ref{Jr}). Indeed,
let us make a substitution $\  r\ \longrightarrow\ \r= r\hbar=\frac{u}{\gamma} $:
\be
[r]=\frac{2\sinh\r}{\{q\}},
\ \ \ \ \ \ \ \{q\}[r-1] = 2\sinh(\rho-\hbar)\ \ \ \ {\rm etc}
\ee
\vspace{-0.2cm}
Then, from (\ref{Jr})
\be
\frac{1}{2}\{q\}{\cal J}^{(k)}(\rho,\hbar) = \sinh\r\,\Big(1 + 4F_1^{(k)}\sinh(\rho-\hbar)\sinh(\rho+\hbar)+\\
+ 16F_2^{(k)}\sinh(\rho-2\hbar)\sinh(\rho-\hbar)\sinh(\rho+\hbar)\sinh(\rho+2\hbar) + \ldots\Big)
=\\=
\sinh\r \sum_{s=0}^\infty (-4k\sinh^2\r)^s\left\{1 + \hbar(k+1)
\left(\frac{2s(s+2)}{3} - \frac{s(s-1)}{6k}\right) + O(\hbar^2)\right\}
= \\
= \frac{\sinh\r}{1+4k\sinh^2\!\r} \
- \ 8k(k+1)\hbar\left(1+\frac{2(2k+1)}{3}\sinh^2\!\r\right)\left(\frac{\sinh\r}{1+4k\sinh^2\!\r}\right)^3
+ O(\hbar^2)
\label{expanJ}
\ee
Similarly, the $\hbar^2$-correction is given by
\be
\!\!\!\!
\frac{1}{30}\sum_{s=0}^\infty \big(2\sinh(\r)\big)^{2s+1}(-k)^{s-2}
\Big(\frac{1}{24}s(s-1)(s-2)(5s+1) - \frac{5}{4}s^2(s+3)(s-1)k
+\frac{5}{24}s(s^3+34s^2+71s-10)k^2 +
\\
+ 5s^2(s+2)(s+3)k^3 +
\frac{2}{3}s(s+3)(5s^2+9s+1)k^4\Big)
- \frac{1}{3}\sum_{s=0}^\infty s(s+1)(2s+1) \big(2\sinh(\r)\big)^{2s-1}(-k)^s =
\\
= \frac{k\sinh\r}{15\Big(1+4k\sinh^2\!\r\Big)^5}\cdot\Big(
(4k^4+10k^3+ 10k^2+5k+1)kx^8+2(k+1)(12k^3+13k^2+7k-2)x^6+
\nn
\ee
\vspace{-0.5cm}
\be
\ \ \ \ \ \ \ \  \ \ \ \ \ \ \ \ \ \ \ \ \ \ \ \ \
+5(2k-1)(6k^2+7k+5)x^4-20(k+1)(2k+1)x^2+60 \Big)
\ee
where $x= 2\sinh\r$.
We do not write down the next corrections, since they are too long.


\newpage
\subsubsection{Measure from the Laplace evolution}

Now, applying (\ref{LevJones}), one obtains:
\be
\mu_\gamma^{(k)} =
\frac{\sinh\left(\frac{u}{\gamma}\right)}{1+4k\sinh^2\left(\frac{u}{\gamma}\right)}-  \\
-\frac{\gamma\hbar}{2} \frac{\p^2}{\p u^2}
\frac{\sinh\left(\frac{u}{\gamma}\right)}{1+4k\sinh^2\left(\frac{u}{\gamma}\right)}
- 8k(k+1)\hbar \left(1+\frac{2(2k+1)}{3}\sinh^2\left(\frac{u}{\gamma}\right)\right)
\left(\frac{\sinh\left(\frac{u}{\gamma}\right)}{1+4k\sinh^2\left(\frac{u}{\gamma}\right)}\right)^3
+ O(\hbar^2) =
\\
= \left(1-\frac{(1-24k)\hbar}{2\gamma}+ O(\hbar^2)\right)\left\{
\frac{\sinh\left(\frac{u}{\gamma}\right)}{1+4k\sinh^2\left(\frac{u}{\gamma}\right)}
\ -\  \hbar\left(\frac{\sinh\left(\frac{u}{\gamma}\right)}{1+4k\sinh^2\left(\frac{u}{\gamma}\right)}\right)^3
\times \right. \\ \left.  \times
\left( \left( \frac{16k(k+1)(2k+1)}{3}+\frac{192k^3}{\gamma}\right)  \sinh\left(\frac{u}{\gamma}\right)^2
 +  \left(8k(k+1)+\frac{16k(7k-1) }{\gamma} \right)\right) + O(\hbar^2)\right\}
\label{muk}
\ee
We definitely calculated a lot more corrections which we used to make our checks.


\subsubsection{The case of torus knots ($k=1$)}

Both coefficients in the last line of (\ref{muk}) vanish in the case of trefoil: for $k=1$ and $\gamma=-6$.
In fact, this remains true for all higher $\hbar$-corrections:
{\bf in the case of trefoil and of other torus knots/links
the evolution operator $e^{-\frac{\gamma\hbar}{2}\p^2_\rho}$ (\ref{LevJones}) converts their Jones
polynomials (expressed via $\r$-variable) into $\hbar$-independent quantities}
(modulo overall $u$-independent normalization factor).

As already mentioned we use the trefoil to illustrate the {\it generic} feature
of the torus family.
The peculiarities are two: for $k=1$ the leading-order measure can be rewritten
in the form of (\ref{BEM}):
\be
\hbar {\cal J}^{(1)} = \frac{\sinh\r}{1+4\sinh^2\!\r} + O(\hbar)
\label{J31large}
\ee
and
\be
\boxed{
 \frac{\sinh\r}{1+4\sinh^2\!\r}
= \frac{\sinh(2\r)\sinh(3\r)}{\sinh(6\r)}
}
\label{torid}
\ee
Moreover, this answer is exact: all $\hbar$-corrections to (\ref{J31large})
are {\it exactly} eliminated by the action of Laplace exponential.
This latter property is true only for $k=1$,
and also depends on the clever choice of $\gamma=-mn=-6$.
More accurately, for such $\gamma$
all the corrections can be absorbed into overall
$u$-independent normalization coefficient in front of $\mu(u)$,
which drops away from the ratio of integrals and do not affect
the averages.

Extension of this result to other torus knots is not at all trivial.
Already for the next $2$-strand knot $[2,5]=5_1$ the relevant analogue of
identity (\ref{torid}) is
\be
 \frac{\sinh(2\rho)\sinh(5\rho)}{\sinh(10\rho)} = \frac{\sinh\r}{1+12\sinh^2\!\r
+ 16\sinh^4\!\r} = {x\over 2}\Big(1 - 3\cdot x^2 + 8\cdot x^4 - 21\cdot x^6 +\ldots\Big)
\ee
with $x=2\sinh\rho$ and to obtain the r.h.s. from Jones polynomial one needs to know the large-$r$
asymptotics of the coefficients $g_{r,j}$ in eq.(64) of \cite{Arth}, e.g.
\be
g_{r,1}^{[2,5]}\sim  3\cdot r, \ \ \ \ldots
\ee
Again after that one can adjust $\gamma=-10$ so that all $\hbar$-corrections
are eliminated by the action of exponentiated Laplacian.


\subsubsection{Checks for $N=2$}

Despite we derived the (formal series for) the measure, starting
from Jones polynomial in the variables $(\r,\hbar)$,
we obtained the answer, which can be used in (\ref{Hasav}) to
evaluate HOMFLY polynomials in concrete representations $[r-1]$,
starting from the fundamental one.

At $N=2$ we can calculate Jones polynomials, but in variables $(r,\hbar)$,
i.e. check that (\ref{Hasav}) reproduces (\ref{Jr}).
More precisely, this works up to a factor:
what is reproduced is the series (\ref{Jr}) times $q^{-\alpha(r^2-1)}$.
Comparing this factor with that in front of the integral transform
(\ref{intra}), it is easy to anticipate $\alpha = -\frac{1}{2}\gamma$.

The measure $\Big<\ldots\Big>^{(k)}$ which reproduces (\ref{Jr}) should satisfy
\be
q^{-\alpha(r^2-1)} J_{r-1}^{(k)}(q) = \left< \frac{\sinh(ru)}{\sinh(u)}\right>^{(k)}
= \sum_{p=0}^\infty c_{r,p} \left< u^{2p} \right>^{(k)}
\label{cons}
\ee
and the claim is that it is indeed given by (\ref{Hasav}) and (\ref{muk}).
The coefficients in (\ref{cons}) are
\be
c_{r,0} = r,  \\
c_{r,1} = \frac{1}{6}r(r^2-1),  \\
c_{r,2} = \frac{1}{360}r(r^2-1)(3r^2-7),  \\
c_{r,3} = \frac{1}{15120}r(r^2-1)(3r^4-18r^2+31),  \\
c_{r,4} = \frac{1}{1814400}r(r^2-1)(5r^6-55r^4+239r^2-381),  \\
\ldots
\ee

The measure and the averages depend on $q=e^\hbar$, and we expand them in series in $\hbar$:
\be
\left< u^{2p} \right>^{(k)} = \sum_{j=0}^\infty \gamma_{p,j}^{(k)}\hbar^{p+j}
\ee
Then we compare the expansion
$
J_r^{(k)} = \sum_{i=0}^\infty j_{r,i}^{(k)} \,\hbar^i,
$
obtained directly from (\ref{Jr}), with
$
\sum_{p,j}^\infty c_{r,p} \gamma_{p,j}^{(k)} \,\hbar^{p+j}.
$
Despite the latter one is the double sum, one can use $r$-dependence
to extract from this comparison all the coefficients $\gamma_{p,j}^{(k)}$.
This gets clear from looking at the first terms of the double expansion:
\be
0 = \sum_{p,j}^\infty c_{r,p} \gamma_{p,j}^{(k)} \, \hbar^{p+j} - J_r^{(k)} =  \\
= \ r\Big(\gamma_{0,0}^{(k)} - 1\Big)  \
+ \ \frac{1}{6}\hbar r^3\Big(\gamma_{1,0}^{(k)} - 6\alpha\Big)
\ +\  \hbar r\Big(\gamma_{0,1}^{(k)} - \frac{1}{6}\gamma_{1,0}^{(k)} + \alpha\Big)
\ +\  \frac{1}{120}\hbar^2 r^5\Big(\gamma_{2,0}^{(k)} - 60\alpha^2\Big) +  \\
+ \ \frac{1}{6}\hbar^2 r^3\Big(\gamma_{1,1}^{(k)}-\frac{1}{6}\gamma_{2,0}^{(k)} - 1+24k+6\alpha^2\Big)
\ +\  \hbar^2 r\Big(\gamma_{0,2}^{(k)}-\frac{1}{6}\gamma_{1,1}^{(k)}+\frac{7}{360}\gamma_{2,0}^{(k)}
\ +\ \frac{1}{6}-4k- \frac{1}{2}\alpha^2\Big) \ +\ \ldots
\ee
The first item at each bracket is expressed through the others, already determined at the
previous stage.
In this way we obtain:
\be
\gamma_{p,0}^{(k)} = \frac{(4\alpha)^p\,\Gamma(p+\frac{3}{2})}{\Gamma(\frac{3}{2})}   \\
\gamma_{p,1}^{(k)} = \frac{(1-24k+4\alpha^2)\,p}{24\alpha^2} \cdot
\frac{(4\alpha)^{p+1}\,\Gamma(p+\frac{3}{2})}{\Gamma(\frac{3}{2})}
\ee
\be
\gamma_{p,2}^{(k)} =
\frac{3(1-240k+1920k^2\, + \, 16\alpha^4)p(p-1)  + 40(1-24k)\alpha^2p(p+1) +2880k(k+1)\alpha p }
{96\cdot 60\,\alpha^4}
\cdot
\frac{(4\alpha)^{p+2}\,\Gamma(p+\frac{3}{2})}{\Gamma(\frac{3}{2})}
\nn
\ee
\be
\!\!\!\!\!\!\!\!\!\!\!\!\!\!\!\!\!\!\!\!\!\!\!\!\!\!\!\!\!\!
\gamma_{p,3}^{(k)} = \Big((1-2184k+67200k^2-322560k^3   + 64\alpha^6)p(p-1)(p-2)
+ 28\Big((1-240k+1920k^2)\alpha^2+4(1-24k)\alpha^4\Big)(p+2)p(p-1) \ +  \\
\!\!\!\!\!\!\!\!\!\!\!\!\!\!\!\!\!\!\!\!\!\!\!
+\ 6720k(k+1)(64k-7)\alpha p(p-1)-26880k(k+1)\alpha^3p(p+1) - 53760k(k^2+4k+1)\alpha^2p\Big)
\cdot\frac{1}{384\cdot 840\,\alpha^6}\cdot
\frac{(4\alpha)^{p+3}\,\Gamma(p+\frac{3}{2})}{\Gamma(\frac{3}{2})}  \\
\ldots
\nn
\ee
\bigskip

\noindent
With our formulas one can check that these parameters are indeed reproduced by (\ref{Hasav}),
to the accuracy of the first three orders of $\hbar$-expansion.

\bigskip

\subsubsection{The cases of $N=3$ and $N=4$}

The same check can be performed for higher $N>2$,
making use  of the measure (\ref{aveL}).
This time HOMFLY polynomials could be reproduced
only if $\tau$ in (\ref{aveL}) is not just $\hbar$,
but rather $\frac{N\hbar}{2}$. We now provide a little more details.

Let us fix an anzatz for $\tau=T\hbar$, where $T$ is a constant that we are going to determine, i.e.
$T$ is the coefficient in front of the $\hbar$-correction to the measure:
\be
\prod_{1\leq i<j\leq N}\sinh^2(u_{ij})\Big(\mu_{(0)}(u_{ij}) + T\hbar \mu_{(1)}(u_{ij})+T^2\hbar^2 \mu_{(2)}(u_{ij})
+ O (\hbar^3)\Big)
\label{mesN}
\ee
with
\be
\mu_{(0)} = \frac{\sinh\left(\frac{u}{\gamma}\right)}{
\sinh(u)\left(1+4k\sinh^2\left(\frac{u}{\gamma}\right)\right)}
\ee
etc. The same measure (\ref{mesN}) is used in the numerator and denominator.

The ratios of matrix model integrals and the corresponding HOMFLY polynomials ${\mathfrak{R}}$
in the order $\hbar^3$ for the fundamental representation ($r=2$) is (up to a power of $q$)
\be
{\mathfrak{R}}\sim 1+24\hbar^3(T-1)f(k,\gamma)
\ee
for $SU(2)$, and
\be
{\mathfrak{R}}\sim 1+32\hbar^3(2T-3)f(k,\gamma)
\ee
for $SU(3)$.  For $N=4$ the factor is $T-2$, so in general it is probably $T-\frac{1}{2}N$.

The values of ${\mathfrak{R}}$ at $A=q^3$ and $A=q^4$ in representations
$[r-1]=[1]$ and $[2]$ are equal to:

\be
\begin{array}{ccc}
N=2,\  [1]: & q^{2\gamma} &
1-\frac{24k\hbar^3}{\gamma}\Big(14k-2+\gamma+\gamma k\Big)(T-1)+O(h^4)
\\
N=2,\  [2]: & q^{6\gamma} &
1-\frac{64k\hbar^3}{\gamma}\Big(14k-2+\gamma+\gamma k\Big)(T-1)+O(h^4)
\\
\hline
\\
N=3,\  [1]: & q^{3\gamma} &
1-\frac{64k\hbar^3}{\gamma}\Big(14k-2+\gamma+\gamma k\Big)\left(T-\frac{3}{2}\right)+O(h^4)
\\
N=3,\  [2]: &q^{8\gamma} &
1-\frac{160k\hbar^3}{\gamma}\Big(14k-2+\gamma+\gamma k\Big)\left(T-\frac{3}{2}\right)+O(h^4)
\\
\hline
\\
N=4,\  [1]: & q^{4 \gamma}&
1-\frac{120k\hbar^3}{\gamma}\Big(14k-2+\gamma+\gamma k\Big)(T-2)+O(h^4)
\\
N=4,\  [2]: & q^{10 \gamma}&
1-\frac{288k\hbar^3}{\gamma}\Big(14k-2+\gamma+\gamma k\Big)(T-2)+O(h^4)
\\
&\ldots&
\end{array}
\ee
Here in the second column we put the normalization factor that differs the matrix model integral and the HOMFLY polynomial in the
topological framing.

Thus, the answer looks like
\be
\boxed{{\mathfrak{R}}=q^{r(N+r-1)\gamma}
\left[
1 -
\frac{8r(N-1)(N+r)k\hbar^3}{\gamma}\Big(14k-2+\gamma+\gamma k\Big)\left(T-\frac{N}{2}\right)+O(h^4)
\right] }
\ee
Since the coefficient in the brackets vanishes for
\be
\gamma\ \stackrel{?}{=}\ -\frac{2\,(7k-1)}{k+1}
\ee
one may think that there is no need to require $T={N\over 2}$
(though for the figure eight knot this does not work anyway).
However, this does not work already in the next order $\hbar^4$:
for generic $k$ the correction does not vanish for this choice of $\gamma$. The only exception is the
case $k=1$ (trefoil): then for this choice ($\gamma=-6$) the corrections do vanish at all orders in $\hbar$.

In the cases of higher $N>2$ one can also consider more complicated representations than just symmetric ones. For instance, for
$N=3$ one can compare the result of matrix model calculation with the HOMFLY polynomial at $A=q^3$ in representation $[21]$
which is known for the trefoil and for the figure eight. We assume that, at least, in the leading orders the HOMFLY polynomial
for other twist knots is given by the same functions $F^{(k)}_p$. Then, the result in this case reads
\be
{\mathfrak{R}}=q^{9\gamma}\left(1-{72k(14k-2+\gamma+\gamma k)(2T-3)\over\gamma}\hbar^3\right.-
\ee
\be
\left.
-24k(2T-3)\left[(2k^2+15k+1)(2T+3)+{8T(k+1)(31k-4)
\over\gamma}+{(828k^2-186k+6)(2T-3)\over\gamma^2}\right]\hbar^4+O(\hbar^5)\right)
\nn
\ee
and one can conclude that the formulas are still correct in this case, supporting the idea
to use the same $F^{(k)}_p$ for all representations
(see sec.4.3 of ref.\cite{Arth} for more careful formulation of this hypothesis).
This example is just the simplest illustration of power of the matrix model approach:
even when being not completed, it already provides new results, which are very difficult to
get by other methods.

\subsubsection{The $\hbar^5$-corrections: violation of universality}

Unfortunately, in higher orders of $\hbar$ the described procedure does not give a complete answer for $k\ne 1$: starting from the order $\hbar^5$ one needs to correct the matrix model result in order to reproduce the right value of the knot polynomial. These corrections
can be absorbed into the normalization factor in (\ref{ga}): they have a rather simple dependence on the representation and $N$.
For instance, in the order $\hbar^5$ they are:
\be\label{h5}
{\mathfrak{R}}=q^{r(N+r-1)\gamma}\left[
1+2(3N-4)(N-2)\Big({2N\varkappa_R}+{r(N^2-r)}\Big)U(k,\gamma){\hbar^5\over\gamma^3}+O(\hbar^6)\right]
\ee
where the quantity
\be
U(k,\gamma)=k\Big(4\gamma^3k^2+3\gamma^3k-\gamma^3+1872k^3-624k^2+48k\Big)
\ee
does not depend on $R$ and $N$, and at $k=1$ factorizes as
\be
U(1,\gamma)=6(\gamma+6)(\gamma^2-6\gamma+36){\hbar^5\over\gamma^3}
\ee
Thus, as before, the correct behaviour of the trefoil is guaranteed at $\gamma=-6$ in higher orders as well.

The coefficient $\varkappa_R$ in (\ref{h5}) is the eigenvalue of the second Casimir operator:
\be
\varkappa_R=\sum_{i,j\in R}(j-i)
\ee
where $R$ denotes the Young diagram corresponding to the representation $R$ and the sum goes over the boxes of this Young
diagram with coordinates $(i,j)$. This quantity is also the eigenvalue of the simplest cut-and-join operator $\widehat W_{[2]}$
\cite{GD}. One can expect that the higher orders in $\hbar$ could be described by higher Casimir or cut-and-join operators \cite{MMN}
in the spirit of \cite{MMS}.

\bigskip

\subsection{Conclusion}

In this letter we developed a systematic approach to the study of
the TBEM-like integral (matrix model) representations of knot polynomials.

\bigskip

The starting point is the Jones polynomial as a function of representation
variable: then the action of exponentiated Laplace operator immediately
provides a measure for an integral representation of the original Jones polynomial --
which in the case of the trefoil is exactly the right TBEM measure for
$N=2$. This, however, is true only if the evolution "time" is appropriately
adjusted ($\gamma=-6$).
Two immediate questions here are: how this works for other torus knots,
and what happens, if one deforms the Laplace evolution.

The next step is lifting the measure from $N=2$ to higher $N$.
The natural prescription (\ref{ga}) is in fact equivalent to
promoting the 2-particle Calogero-Ruijsenaars evolution to the $N$-particle one.
Again, a natural question is what happens, if one allows higher
Hamiltonians to contribute.

We demonstrated that all these questions can indeed be relevant,
because the above two-step procedure works perfectly only for the trefoil:
the Laplace evolution and its ordinary lifting to higher $N$ is indeed
equivalent to the TBEM model (though an exact proof is still needed even
in this case).

Already for the family of twist knots there are corrections
which clearly exhibit a clever representation dependence to be described
by some kind of a deformation of the Laplace evolution (by higher Hamiltonians,
i.e. Casimirs, or, perhaps, by more general cut-and-join operators \cite{MMN}).
Unfortunately, the technique of $\hbar$-expansion which we used in this letter
(and in a closely related investigation of the Hurwitz $\tau$-function structure
of colored HOMFLY polynomials and superpolynomials in \cite{MMS})
is not sufficient to answer these questions.

\bigskip

More powerful methods of group, matrix model, conformal and integrability theories
should now be applied to attack this very promising problem.
Among immediate topics to study are the clearly seen relations to
the volume conjecture and Hikami invariants \cite{Kash,Hik,VC}.

\section*{Acknowledgements}

We are indebted to our numerous colleagues, who participated at
different stages of our search of the matrix model for HOMFLY
polynomials: P.Dunin-Barkowski, D.Galakhov, D.Melnikov, V.Pestun,
A.Popolitov, A.Sleptsov, A.Smirnov, Sh.Shakirov.
Our work is partly supported by ERC Starting Independent Researcher Grant StG No. 204757-TQFT (A.A.), by grant
NSh-1500.2014.2,
by RFBR  grants 13-02-00457 (A.A., A.Mir.), 13-02-00478 (A.Mor.), 14-02-00627 (And.Mor.),
by joint grants 13-02-91371-ST, 14-01-92691-Ind,
by the Brazil National Counsel of Scientific and
Technological Development (A.Mor.),
by the Laboratory of Quantum Topology of
Chelyabinsk State University (Russian Federation government grant 14.Z50.31.0020) (And.Mor.) and
by D.~Zimin's ``Dynasty'' foundation (And.Mor.).

\end{document}